\documentclass[11pt]{revtex4}
\usepackage{amssymb}
\usepackage{latexsym}
\usepackage{epsfig}                               

\begin{document}
\title{Interacting Ghost Dark Energy in Non-Flat Universe}

\author{A. Sheykhi $^{1,2}$ \footnote{
sheykhi@mail.uk.ac.ir} and M. Sadegh Movahed$^{3}$ \footnote{
m.s.movahed@ipm.ir} }
\address{$^1$Department of Physics, Shahid Bahonar University, PO Box 76175, Kerman, Iran\\
         $^2$Research Institute for Astronomy and Astrophysics of Maragha (RIAAM), Maragha,
         Iran\\ $^3$  Department of Physics, Shahid Beheshti university, G.C., Evin, Tehran 19839, Iran
}

\begin{abstract}
A new dark energy model called `` ghost dark energy" was recently
suggested  to explain the observed accelerating expansion of the
universe. This model originates from the Veneziano ghost of QCD. The
dark energy density is proportional to Hubble parameter,
$\rho_D=\alpha H$, where $\alpha$ is a constant of order
$\Lambda_{\rm QCD}^3$ and $\Lambda_{\rm QCD}\sim 100 MeV$ is QCD
mass scale. In this paper, we extend the ghost dark energy model to
the universe with spatial curvature in the presence of interaction
between dark matter and dark energy. We study cosmological
implications of this model in detail. In the absence of interaction
the equation of state parameter of ghost dark energy is always $w_D
> -1 $ and mimics a cosmological constant in the late time, while it
is possible to have $w_D < -1 $ provided the interaction is taken
into account. When $k = 0$, all previous results of ghost dark
energy in flat universe are recovered. To check the observational consistency,
we use Supernova type Ia (SNIa) Gold sample, shift parameter of
Cosmic Microwave Background radiation (CMB) and the Baryonic
Acoustic Oscillation peak from Sloan Digital Sky Survey (SDSS). The best fit values of free parameter at $1\sigma$ confidence interval are:   $\Omega_m^0=
0.35^{+0.02}_{-0.03}$, $\Omega_D^0=0.75_{-0.04}^{+0.01}$ and
$b^2=0.08^{+0.03}_{-0.03}$. Consequently the total energy density of universe at present time in this model at $68\%$ level equates to $\Omega_{\rm
tot}^0=1.10^{+0.02}_{-0.05}$.
\end{abstract}

 \maketitle

\section{Introduction}
The current acceleration of the cosmic expansion has been strongly
confirmed  by numerous and complementary observational data
\cite{Rie}. In the context of standard cosmology such an expansion
requires the existence of an unknown dominant energy component,
usually dubbed ``dark energy'' whose equation of state parameter
satisfies $w_D<-1/3$. Although we can affirm that the ultimate
fate of the universe is determined by the feature of dark energy,
the nature of dark energy as well as its cosmological origin is
still rather uncertain. (for reviews, see e.g. \cite{Lobo} and
references therein). Disclosing the nature of dark energy has been
one of the most important challenges of the modern cosmology and
theoretical physics in the past decade. A great varieties of dark
energy models have been proposed, to explain the acceleration of
the universe expansion within the framework of quantum gravity, by
introducing new degree of freedom or by modifying the underlying
theory of gravity \cite{Li,cop,sheykhi,Cai1}.

Recently a very interesting suggestion on the origin of a dark
energy is made, without introducing new degrees of freedom beyond
what are already known, with the dark energy of just the right
magnitude to give the observed expansion \cite{Urban,Ohta}. In this
proposal, it is claimed that the cosmological constant arises from
the contribution of the ghost fields which are supposed to be
present in the low-energy effective theory of QCD
\cite{Wit,Ven,Ros,Na,Kaw}. It was argued that the Veneziano ghost,
which is unphysical in the usual Minkowski spacetime QFT, exhibits
important physical effects in dynamical spacetime or spacetime with
non-trivial topology. The ghosts are required to exist for the
resolution of the $U(1)$ problem, but are completely decoupled from
the physical sector \cite{Kaw}. The above claim is that the ghosts
are decoupled from the physical states and make no contribution in
the flat Minkowski space, but once they are in the curved space or
time-dependent background, the cancelation of their contribution to
the vacuum energy is off-set, leaving a small energy density
$\rho\sim H \Lambda^3_{QCD}$, where $H$ is the Hubble parameter and
$\Lambda_{QCD}$ is the QCD mass scale of order a $100 MeV$. With
$H\sim 10^{-33} eV$, this gives the right magnitude $\sim (3\times
10^{-3} eV)^4$ for the observed dark energy density. This numerical
coincidence is remarkable and also means that this model gets rid of
fine tuning problem \cite{Urban,Ohta}. The advantages of this new
model compared to other dark energy models is that it is totally
embedded in standard model and general relativity, one needs not to
introduce any new parameter, new degree of freedom or to modify
gravity. The dynamical behavior of the ghost dark energy (GDE) model
in flat universe have been studied \cite{CaiGhost}.

In this paper we would like to extend the previous discussion on
ghost dark energy \cite{CaiGhost} to a universe with spatial
curvature. There are enough observational evidences, at present
time, for taking into account a small but non-negligible spatial
curvature \cite{spe}. For instance, the tendency of preferring a
closed universe appeared in a suite of CMB experiments \cite{Sie}.
The improved precision from WMAP provides further confidence,
showing that a closed universe with positively curved space is
marginally preferred \cite{Uzan}. In addition to CMB, recently the
spatial geometry of the universe was probed by supernova
measurements of the cubic correction to the luminosity distance
\cite{Caldwell}, where a closed universe is also marginally favored.

Most discussions on dark energy models rely on the fact that its
evolution is independent of other matter fields. Given the unknown
nature of both dark matter and dark energy there is nothing in
principle against their mutual interaction and it seems very special
that these two major components in the universe are entirely
independent. Indeed, this possibility has got a lot of attention in
the literature in recent years (see \cite{Ame,Zim,wang1} and
references therein) and was shown to be compatible with SNIa and CMB
data \cite{Oli}.

All above reasons, motivate us to study the interacting ghost dark
energy model in a nonflat universe. In this paper, we would like to
generalize the ghost dark energy model to the universe with spacial
curvature in the presence of interaction between the dark matter and
dark energy. Taking the interaction between the two different
constituents of the universe into account, we study the evolution of
the universe, from early deceleration to late time acceleration. In
addition, we will show that such an interacting dark energy model
can accommodate a transition of the dark energy from a normal state
where $w_D > -1$ to $w_D < -1$ phantom regimes.

This paper is organized as follows. In the next section, we review
the ghost dark energy model in a flat universe. In section
\ref{nonflat}, we generalize the study to the universe with spacial
curvature in the presence of interaction between dark matter and
dark energy. Observational constraints  on the free parameters of model will be given in section IV. We summarize our results in section V.

\section{Ghost dark energy in flat universe}\label{flat}
Let us first review the ghost dark energy model in flat
Friedmann-Robertson-Walker (FRW) universe where first investigated
in \cite{CaiGhost}. Although, our approach in dealing with the
problem differs to some extent from those of Ref. \cite{CaiGhost}.

\subsection{Noninteracting case}
For the flat FRW universe filled with dark energy and dust (dark
matter), the corresponding Friedmann equation takes the form
\begin{eqnarray}\label{Fried}
H^2=\frac{1}{3M_p^2} \left( \rho_m+\rho_D \right),
\end{eqnarray}
where $\rho_m$ and $\rho_D$ are, respectively, the energy
densities of pressureless matter and dark energy. The ghost energy
density is \cite{Ohta}
\begin{equation}\label{GDE}
\rho_D=\alpha H,
\end{equation}
where $\alpha$ is a constant of order $\Lambda_{\rm QCD}^3$ and
$\Lambda_{\rm QCD}$ is QCD mass scale. With $\Lambda_{\rm QCD}\sim
100MeV$ and $H\sim 10^{-33}eV$ ,  $\Lambda_{\rm QCD}^3 H$ gives the
right order of magnitude $\sim (3\times 10^{-3}\rm {eV})^4$ for the
observed dark energy density \cite{Ohta}.

\begin{figure}[htp]
\begin{center}
\includegraphics[width=8cm]{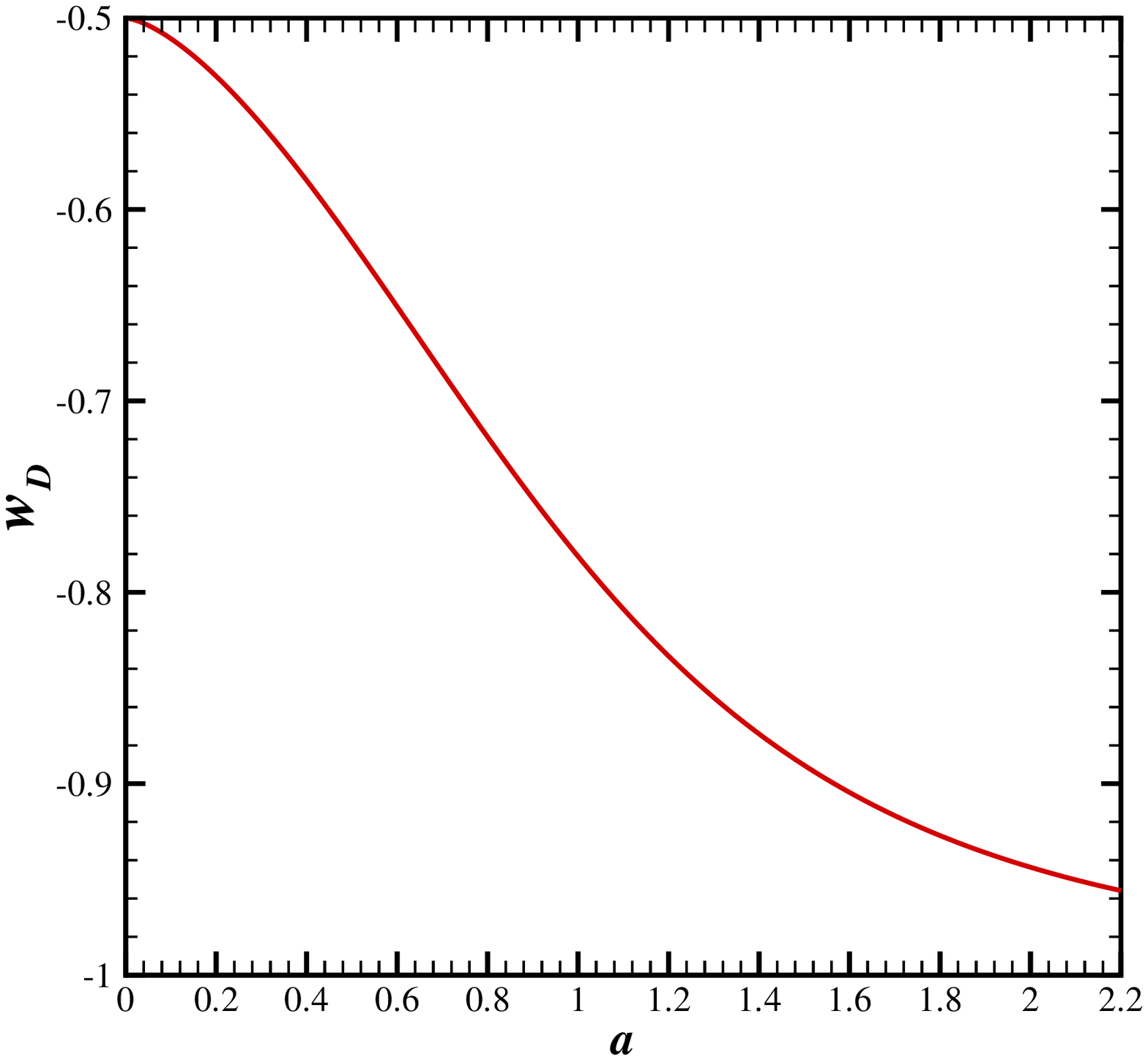}
\includegraphics[width=8cm]{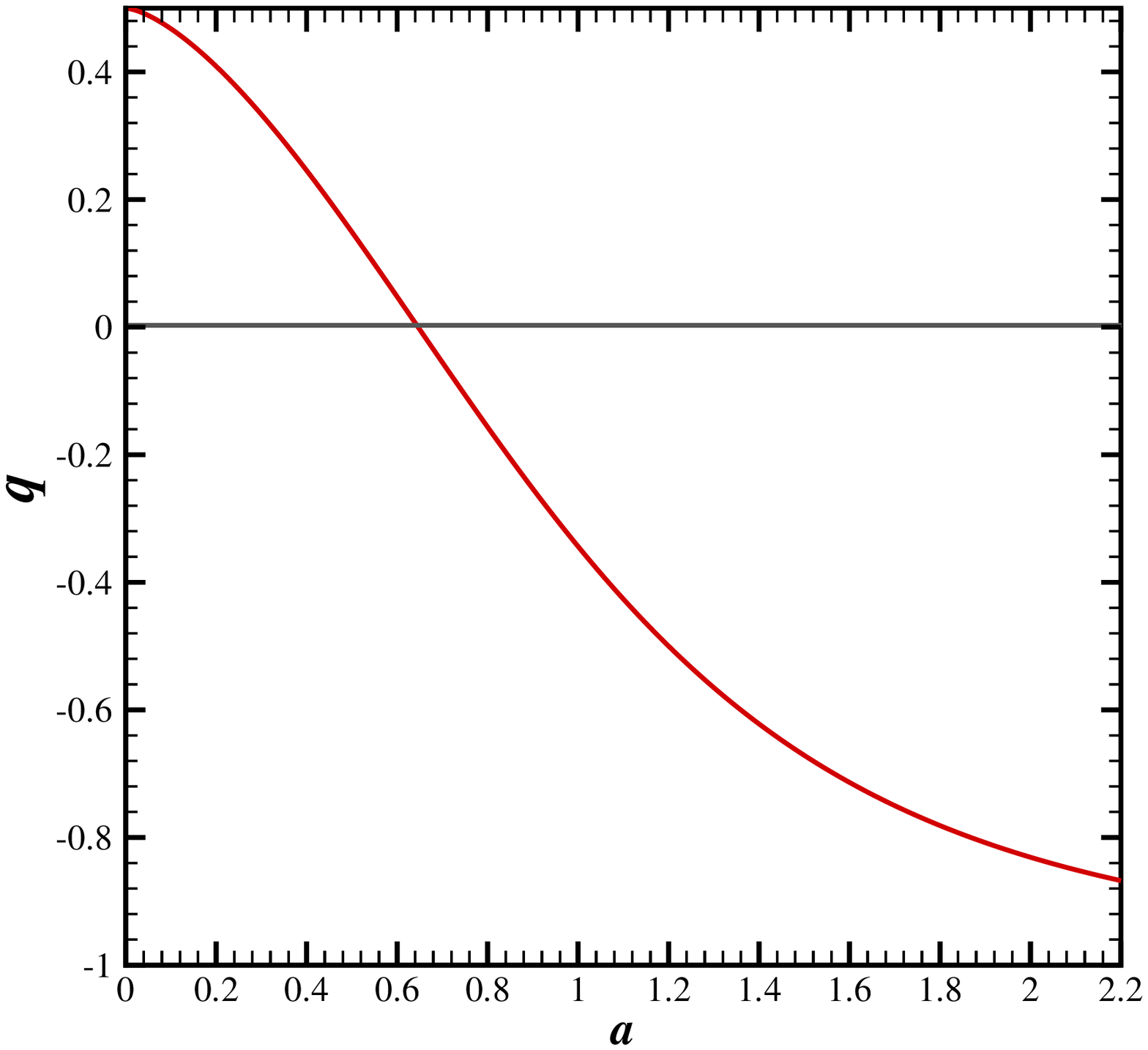}
\caption{Left panel shows the evolution of $w_D$ for ghost dark
energy. In the right panel the behavior of the deceleration
parameter for ghost dark energy is illustrated. Here we have taken
$\Omega^{0}_{D}=0.72.$ }\label{fig1}
\end{center}
\end{figure}

We define the dimensionless density parameters as
\begin{equation}\label{Omega}
\Omega_m=\frac{\rho_m}{\rho_{cr}},\ \ \
\Omega_D=\frac{\rho_D}{\rho_{cr}}=\frac{\alpha}{3M_p^2 H},
\end{equation}
where the critical energy density is $\rho_{cr}={3H^2 M_p^2}$. Thus,
the Friedmann equation can be rewritten as
\begin{equation}\label{fridomega}
\Omega_m+\Omega_D=1.
\end{equation}
The conservation equations read
\begin{eqnarray}
\dot\rho_m+3H\rho_m&=&0,\label{consm}\\
\dot\rho_D+3H\rho_D(1+w_D)&=&0\label{consd}.
\end{eqnarray}
Taking the time derivative of relation (\ref{GDE}) and using the
Friedmann equation we find
\begin{equation}\label{dotrho}
\dot{\rho}_D=\rho_D \frac{\dot{H}}{H}=-\frac{\alpha }{2 M_p^2}
\rho_D(1+u+w_D).
\end{equation}
where $u=\rho_m/\rho_D$ is the energy density ratio. Inserting
this relation in continuity equation (\ref{consd}) we reach
\begin{equation}\label{wD}
(1+w_D)(6M_p^2 H-\alpha)=\alpha u.
\end{equation}
Substituting ghost energy density (\ref{GDE}) in Friedmann
equation (\ref{Fried}) we find
\begin{equation}\label{Fried2}
3M_p^2 H=\alpha(1+u).
\end{equation}
Combining Eq. (\ref{Fried2}) with (\ref{wD}) we reach
\begin{equation}\label{wD2}
w_D=-1+\frac{u}{1+2u}.
\end{equation}
Using the fact that
\begin{equation}\label{u}
u=\frac{\rho_m}{\rho_D}=\frac{\Omega_m}{\Omega_D}=\frac{1-\Omega_D}{\Omega_D},
\end{equation}
we can rewrite Eq. (\ref{wD2}) as
\begin{equation}\label{wD3}
w_D=-\frac{1}{2-\Omega_D},
\end{equation}
It is easy to see that at the early time  where $\Omega_D\ll 1$ we
have $w_D=-1/2$, while at the late time where $\Omega_D\rightarrow
1$ the ghost dark energy mimics a cosmological constant, namely
$w_D= -1$.  It is worthy to note that in $w_D$ of this model, there
is no free parameter. In the left panel of figure (\ref{fig1}) we
plot the evolution of $w_D$ versus scale factor $a$. From this
figure we see that $w_D$ of the ghost dark energy model cannot cross
the phantom divide and the universe has a de Sitter phase at late
time.

\begin{figure}[htp]
\begin{center}
\includegraphics[width=8cm]{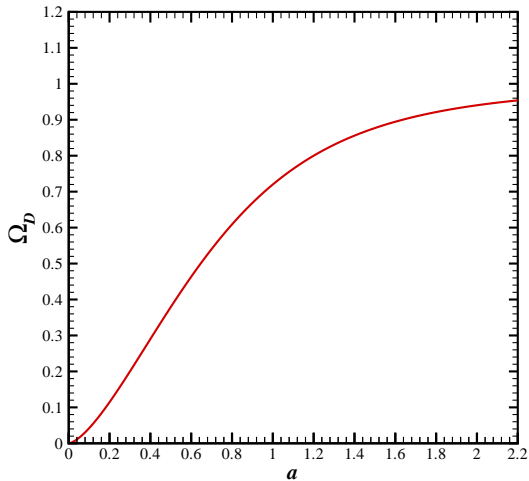}
\caption{The evolution of $\Omega_{D}$ for ghost dark energy. Here
we have taken $\Omega^{0}_{D}=0.72$.}\label{fig2}
\end{center}
\end{figure}

We can also calculate the deceleration parameter which is defined
as
\begin{equation}\label{q1}
q=-1-\frac{\dot{H}}{H^2}.
\end{equation}
When the deceleration parameter is combined with the Hubble
parameter and the dimensionless density parameters form a set of
useful parameters for the description of the astrophysical
observations. Using Eq. (\ref{dotrho}) and definition $\Omega_D$
in (\ref{Omega}) we obtain
\begin{equation}\label{dotH}
\frac{\dot{H}}{H^2}=-\frac{3}{2}\Omega_D \left(1+u+w_D\right).
\end{equation}
Substituting this relation into  (\ref{q1}), after using
(\ref{wD3}) we find
\begin{equation}\label{q2}
q=\frac{1}{2}-\frac{3}{2}\frac{\Omega_D}{(2-\Omega_D)}
\end{equation}
At the early time  where $\Omega_D\rightarrow 0$ the deceleration
parameter becomes $q=1/2$, while at the late time where the dark
energy dominates ($\Omega_D\rightarrow 1$) we have $q=-1$. This
implies that at the early time the universe is in a deceleration
phase while at the late time it enters an acceleration phase.  We
have plotted the behavior of $q$ in the right panel of figure
(\ref{fig1}). From this figure we see that the transition from
deceleration to acceleration take places at $a\simeq0.64$ or
equivalently at redshift $z\simeq 0.56$. Note that $1+z=a^{-1}$ and
we have set $a_0=1$ for the present value of scale factor. Besides,
taking $\Omega_{D}^0=0.72$ we obtain $q \approx-0.34$ for the present
value of the deceleration parameter which is in agreement with
recent observational data \cite{daly}. Taking the time derivative of
Eq. (\ref{Omega}) and using relation ${\dot{\Omega}_D}= H
\frac{d\Omega_D}{d\ln a}$ as well as relation (\ref{q1}) we reach
\begin{equation}\label{Omegaprime1}
\frac{d\Omega_D}{d\ln a}=\Omega_D\left(1+q\right).
\end{equation}
Using Eq. (\ref{q2}) we get
\begin{equation}\label{Omegaprime2}
\frac{d\Omega_D}{d\ln a}=3 \Omega_D
\frac{(1-\Omega_D)}{2-\Omega_D}.
\end{equation}
This is the equation governing the evolution of ghost dark energy.
The dynamics of ghost dark energy is plotted in figure (\ref{fig2})
where we have taken $\Omega^{0}_{D}=0.72$ as the initial condition.
This figure shows that at the late time the dark energy dominates,
as expected.
\subsection{Interacting case}
Next we extend the discussion to the interacting case and study
the dynamics of the ghost dark energy. Although at this point the
interaction may look purely phenomenological but different
Lagrangians have been proposed in support of it (see \cite{Tsu}
and references therein). Besides, in the absence of a symmetry
that forbids the interaction there is nothing, in principle,
against it. In addition, given the unknown nature of both dark
energy and dark matter, which are two major contents of the
universe, one might argue that an entirely independent behavior of
dark energy is very special \cite{wang1,pav1}. Further, the
interacting dark mater-dark energy (the latter in the form of a
quintessence scalar field and the former as fermions whose mass
depends on the scalar field) has been investigated at one quantum
loop with the result that the coupling leaves the dark energy
potential stable if the former is of exponential type but it
renders it unstable otherwise \cite{Dor}. Thus, microphysics seems
to allow enough room for the coupling; however, this point is not
fully settled and should be further investigated. The difficulty
lies, among other things, in that the very nature of both dark
energy and dark matter remains unknown whence the detailed form of
the coupling cannot be elucidated at this stage. In this case, the
energy densities of dark energy and dark matter no longer satisfy
independent conservation laws. They obey instead
\begin{figure}[htp]
\begin{center}
\includegraphics[width=8cm]{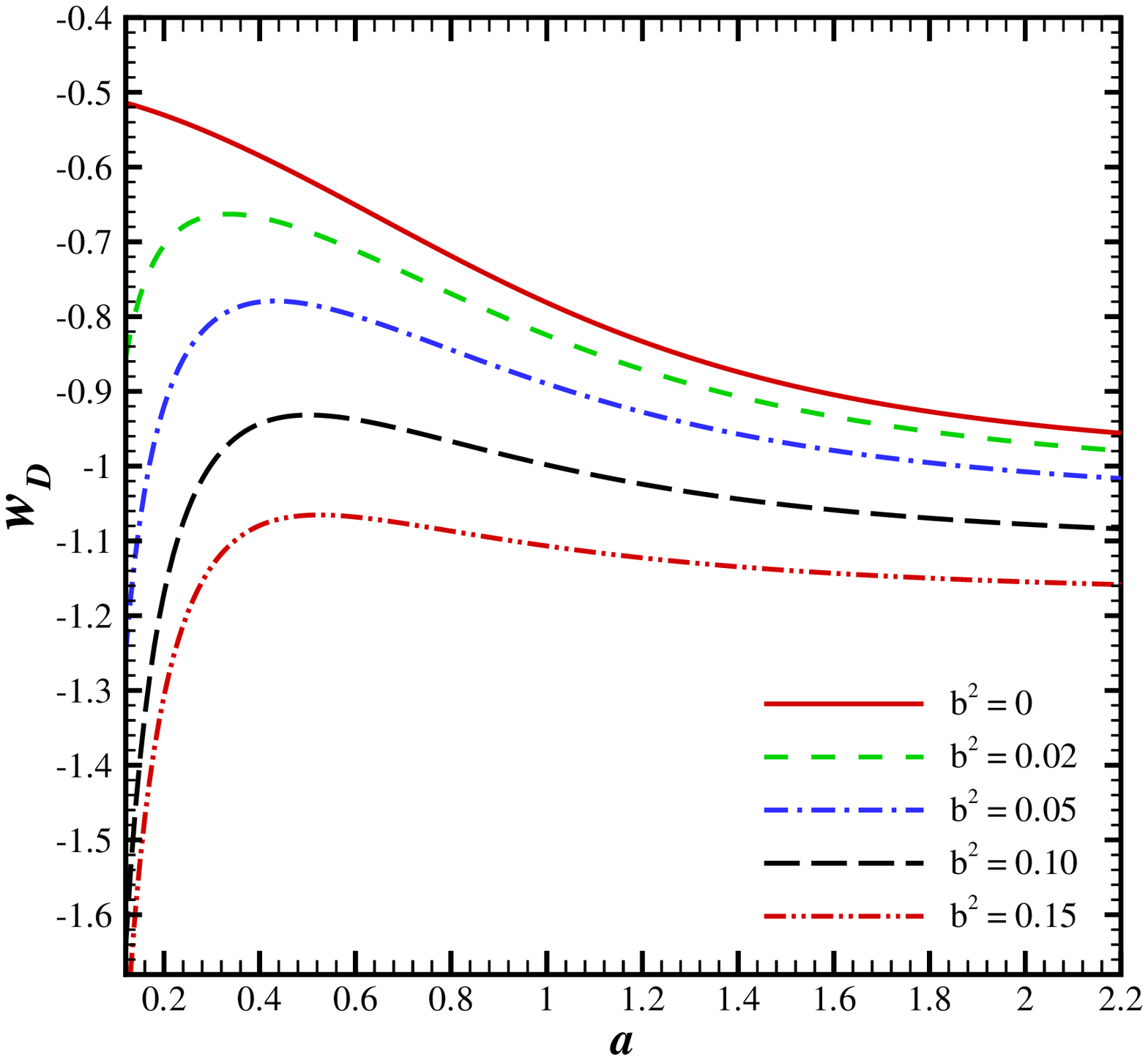}
\includegraphics[width=8cm]{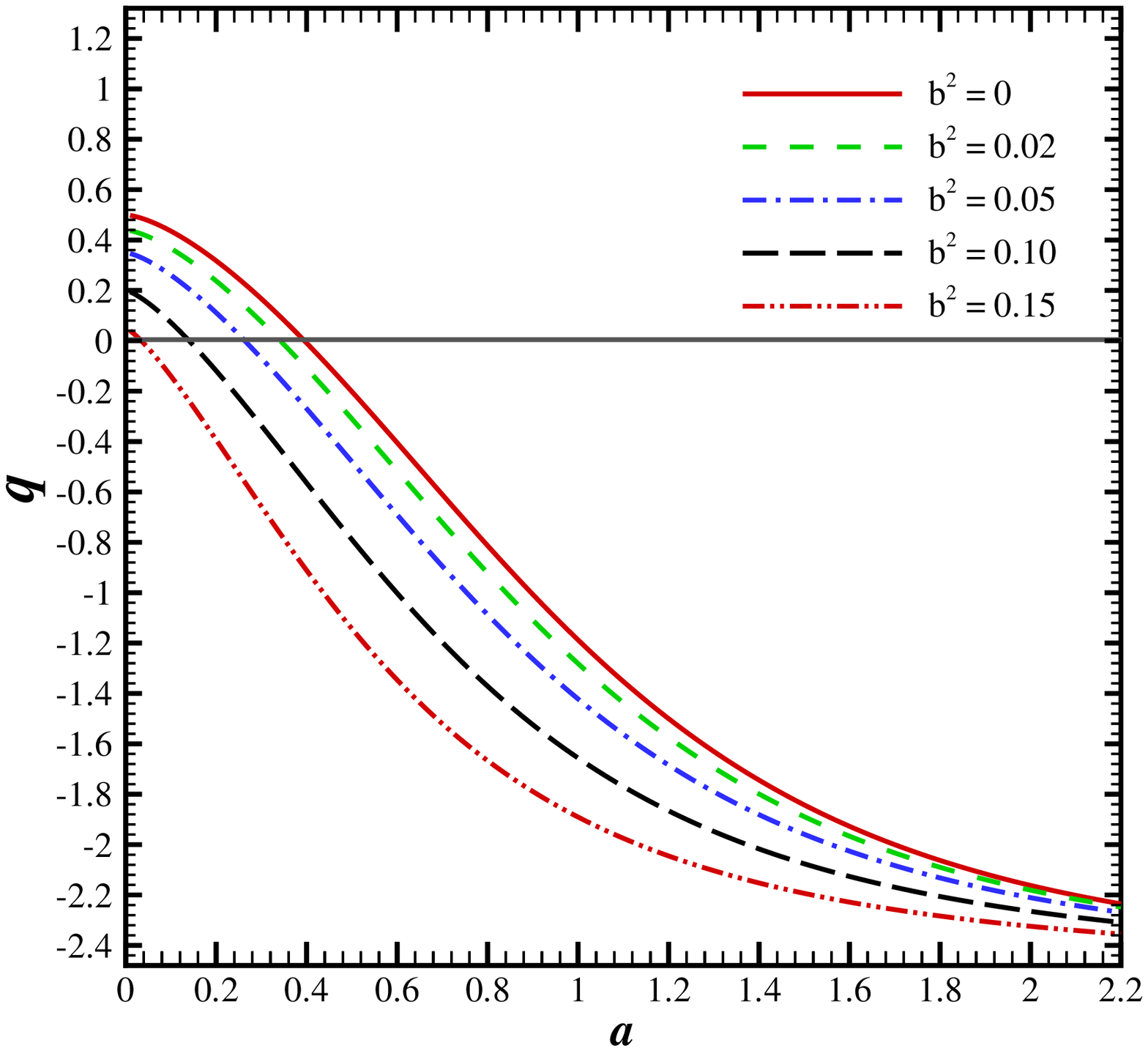}
\caption{Left panel corresponds to the evolution of $w_D$ for
interacting ghost dark energy and different interacting parameter
$b^2$ while right panel shows the evolution of the deceleration
parameter for interacting ghost dark energy and different
interacting parameter $b^2$. Here we took $\Omega^{0}_{D}=0.72.$
}\label{fig4}
\end{center}
\end{figure}
\begin{eqnarray}
\dot\rho_m+3H\rho_m&=&Q,\label{consm2}\\
\dot\rho_D+3H\rho_D(1+w_D)&=&-Q\label{consd2},
\end{eqnarray}
where $Q$ represents the interaction term and we take it as
\begin{equation}\label{Q}
Q =3b^2 H(\rho_m+\rho_D)=3b^2 H\rho_D(1+u),
\end{equation}
with $b^2$  being a coupling constant. It is worth noting that the
continuity equations imply that the interaction term should be a
function of a quantity with units of inverse of time (a first and
natural choice can be the Hubble factor $H$) multiplied with the
energy density. Therefore, the interaction term could be in any of
the following forms: (i) $Q\propto H \rho_D$, (ii) $Q\propto H
\rho_m$, or (iii) $Q\propto H (\rho_m+\rho_D)$. Thus we can present
the above three choices in one expression as $Q =\Gamma\rho_D$,
where
\begin{eqnarray}
\begin{array}{ll}
\Gamma=3b^2H  \hspace{1.3cm}   {\rm for}\  \  Q\propto H \rho_D, &  \\
\Gamma=3b^2Hu \hspace{1.1cm}   {\rm for} \  \ Q\propto H \rho_m,&  \\
\Gamma=3b^2H(1+u) \ \   {\rm for} \  \ Q\propto H (\rho_m+\rho_D),&
  \end{array}
 \end{eqnarray}
It should be noted that the ideal interaction term must be motivated
from the theory of quantum gravity. In the absence of such a theory,
we rely on pure dimensional basis for choosing an interaction $Q$.
In the present work for the sake of generality, we choose the third
expression for the interaction term.

Inserting Eqs. (\ref{dotrho}) and (\ref{Q}) in Eq. (\ref{consd2})
and using (\ref{u}) we find
\begin{equation}\label{wD4}
w_D=-\frac{1}{2-\Omega_D}\left(1+\frac{2b^2}{\Omega_D}\right).
\end{equation}
One can easily check that in the late time where
$\Omega_D\rightarrow 1$, the equation of state parameter of
interacting ghost dark energy necessary crosses the phantom line,
namely, $w_D=-(1+2b^2)<-1$ independent of the value of coupling
constant $b^2$. For present time with taking $\Omega^{0}_D=0.72$,
the phantom crossing can be achieved provided $b^2>0.1$. This value
for coupling constant is consistent with recent observations
\cite{wang1}.
\begin{figure}[htp]
\begin{center}
\includegraphics[width=8cm]{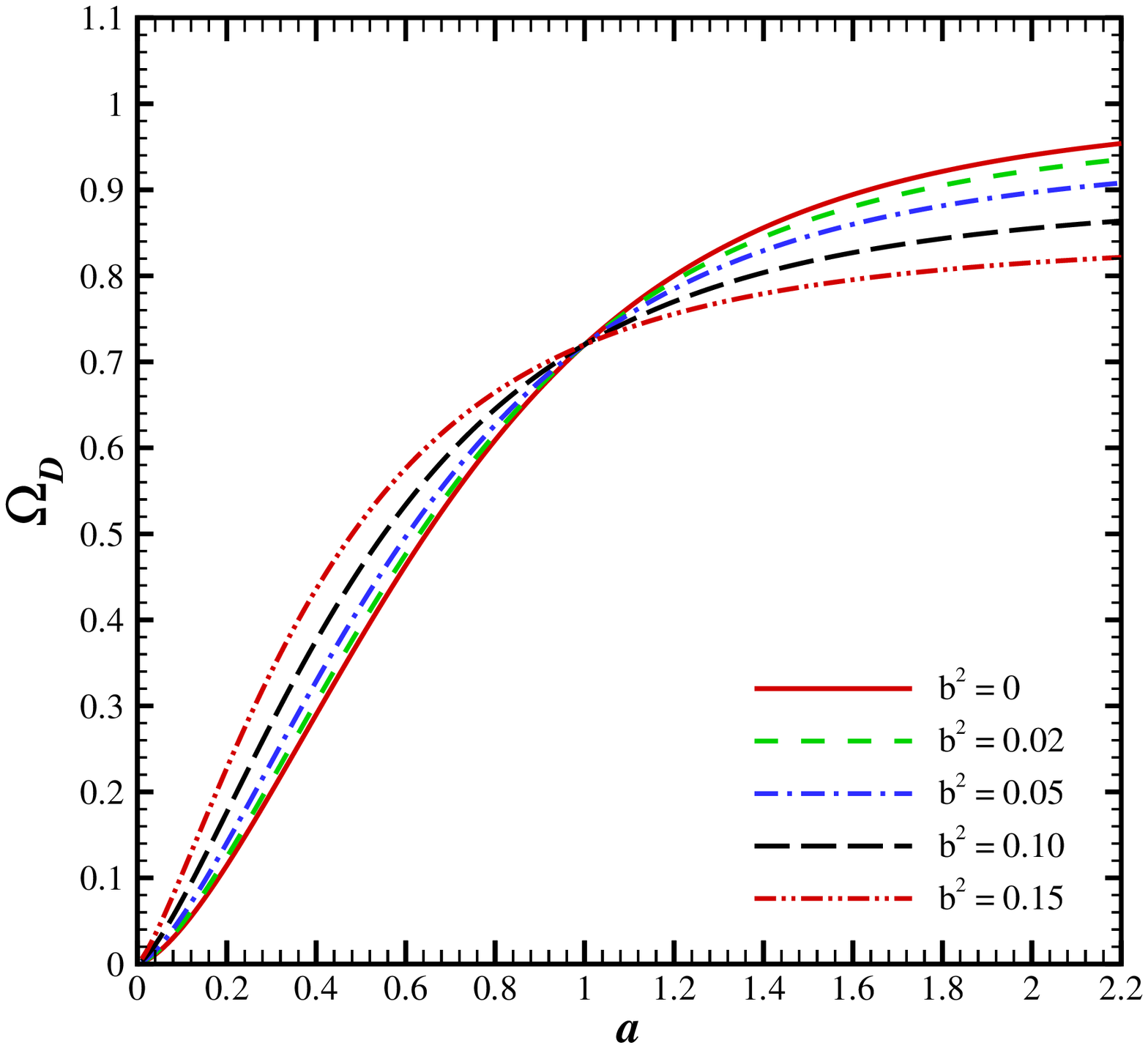}
\includegraphics[width=8cm]{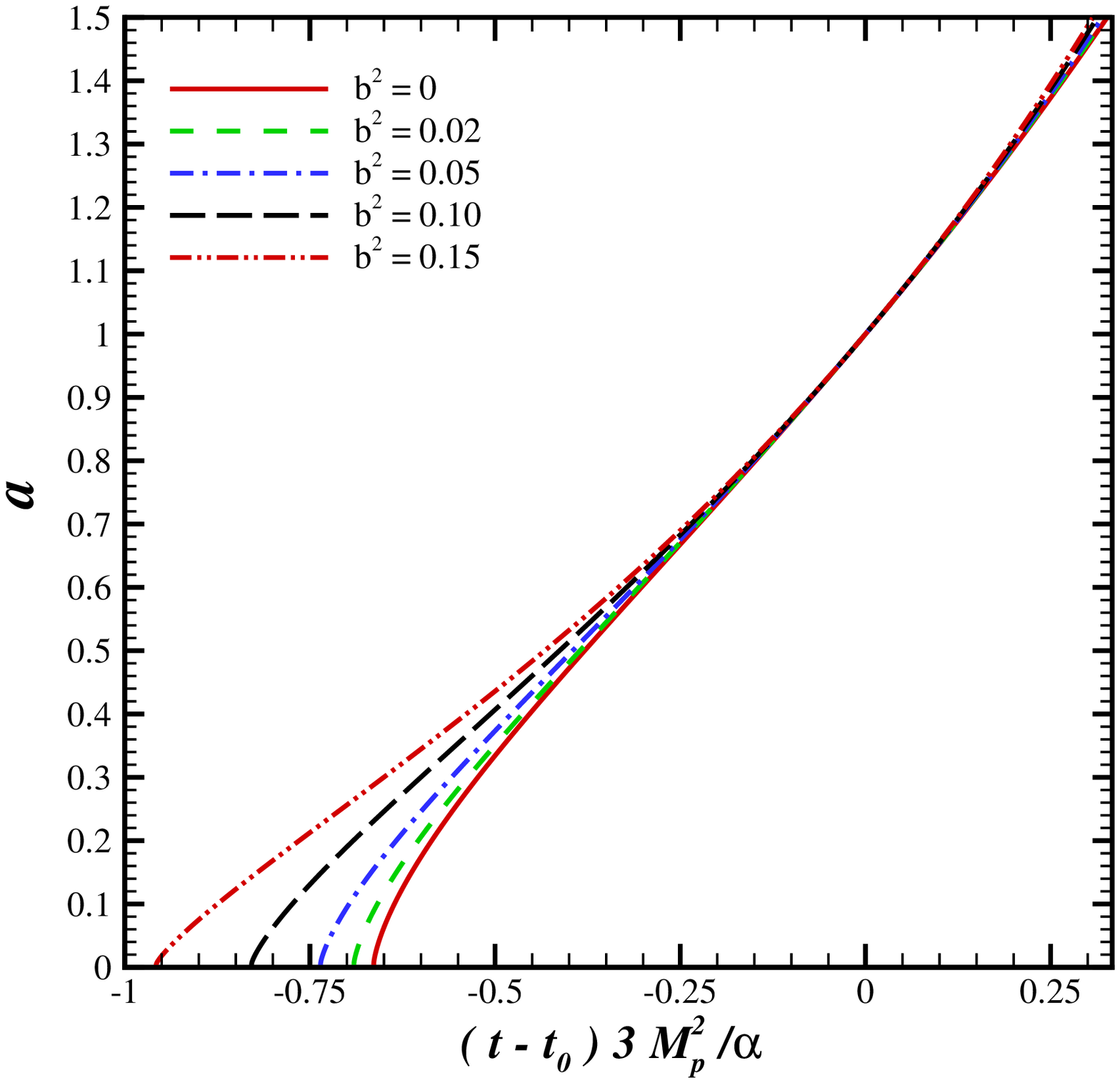}
\caption{The evolution of dark energy density for interacting ghost
dark energy is shown in left panel. Right panel corresponds to the
evolution of the scale factor for interacting ghost dark energy with
different $b^2$. The rest of parameters are the same as for figure
(\ref{fig4}). }\label{fig6}
\end{center}
\end{figure}


In the presence of interaction the deceleration parameter is
obtained by substituting (\ref{wD4}) in (\ref{dotH}) and using
(\ref{q1}). The result is
\begin{equation}\label{q3}
q=\frac{1}{2}-\frac{3}{2}\frac{\Omega_D}{(2-\Omega_D)}\left(1+\frac{2b^2}{\Omega_D}\right),
\end{equation}
while the evolution of dark energy follows the following equation
\begin{equation}\label{Omegaprime3}
\frac{d\Omega_D}{d\ln a}=\frac{3}{2} \Omega_D\left[1-\frac{
\Omega_D}{2- \Omega_D}\left(1+\frac{2b^2}{\Omega_D}\right)\right].
\end{equation}
The evolution of the cosmological parameters $w_D$, $q$ and
$\Omega_D$ are shown in figures  (\ref{fig4}) and (\ref{fig6}) for
different interacting parameter $b^2$. We have taken
$\Omega^{0}_{D}=0.72$ as the initial condition. We can also obtain
the scale factor $a$ as a function of $t$. Integrating the relation
$\Omega_D=\alpha/(3M_p^2 H)$, we find
\begin{equation}\label{at}
\int {\Omega_D
\frac{da}{a}}=\int^{t}_{t_0}{\frac{\alpha}{3M_p^2}dt}=\frac{\alpha}{3M_p^2}(t-t_0),
\end{equation}
where $\Omega_D$ is given by Eq. (\ref{Omegaprime3}). The behavior
of $a(t)$ is shown in the right panel of figure (\ref{fig6}).
\section{Interacting Ghost dark energy in non-flat universe}\label{nonflat}
Next we reach to the main task of the
present work, namely studying the dynamic evolution of ghost
energy density in a universe with special curvature. As we
discussed in the introduction a closed universe is marginally
favored. Taking the curvature into account, the Friedmann equation
is written as
\begin{eqnarray}\label{Friedm}
H^2+\frac{k}{a^2}=\frac{1}{3M_p^2} \left( \rho_m+\rho_D \right),
\end{eqnarray}
where $k$ is the curvature parameter with $k = -1, 0, 1$
corresponding to open, flat, and closed universes, respectively.
We define the curvature density parameter as
$\Omega_k=k/(a^2H^2)$, thus the Friedmann equation takes the form
\begin{equation}\label{fridomega2}
1+\Omega_k=\Omega_m+\Omega_D,
\end{equation}
Using the above equation the energy density ratio becomes
\begin{equation}\label{u2}
u=\frac{\rho_m}{\rho_D}=\frac{\Omega_m}{\Omega_D}=\frac{1+\Omega_k-\Omega_D}{\Omega_D}.
\end{equation}
Taking the time derivative of the Friedmann equation
(\ref{Friedm}) we find
\begin{equation}\label{dotrho2}
\frac{\dot{H}}{H^2}=\Omega_k-\frac{3 }{2} \Omega_D[1+u+w_D],
\end{equation}
and therefore
\begin{equation}\label{dotrho3}
\frac{\dot{\rho}_D}{H}=\rho_D\frac{\dot{H}}{H^2}=\rho_D\left(\Omega_k-\frac{3
}{2} \Omega_D[1+u+w_D]\right).
\end{equation}
Combining this relation with continuity equation (\ref{consd2}),
after using (\ref{Q}) and (\ref{u2})  we find the equation of
state parameter of interacting ghost dark energy in non-flat
universe
\begin{equation}\label{wDn2}
w_D=-\frac{1}{2-\Omega_D}\left(1-\frac{\Omega_k}{3}+\frac{2b^2}{
\Omega_D} (1+\Omega_k)\right).
\end{equation}
The deceleration parameter is obtained as
\begin{equation}\label{q1n}
q=-1-\frac{\dot{H}}{H^2}=-1-\Omega_k+\frac{3 }{2}
\Omega_D[1+u+w_D]
\end{equation}
Substituting Eqs. (\ref{u2}) and (\ref{wDn2}) in (\ref{q1n}) we
obtain
\begin{equation}\label{q2n}
q=\frac{1}{2}\left(1+\Omega_k\right)-\frac{3\Omega_D}{2(2-\Omega_D)}\left[1-\frac{\Omega_k}{3}+2b^2
\Omega_D^{-1} (1+\Omega_k)\right],
\end{equation}
\begin{figure}[htp]
\begin{center}
\includegraphics[width=8cm]{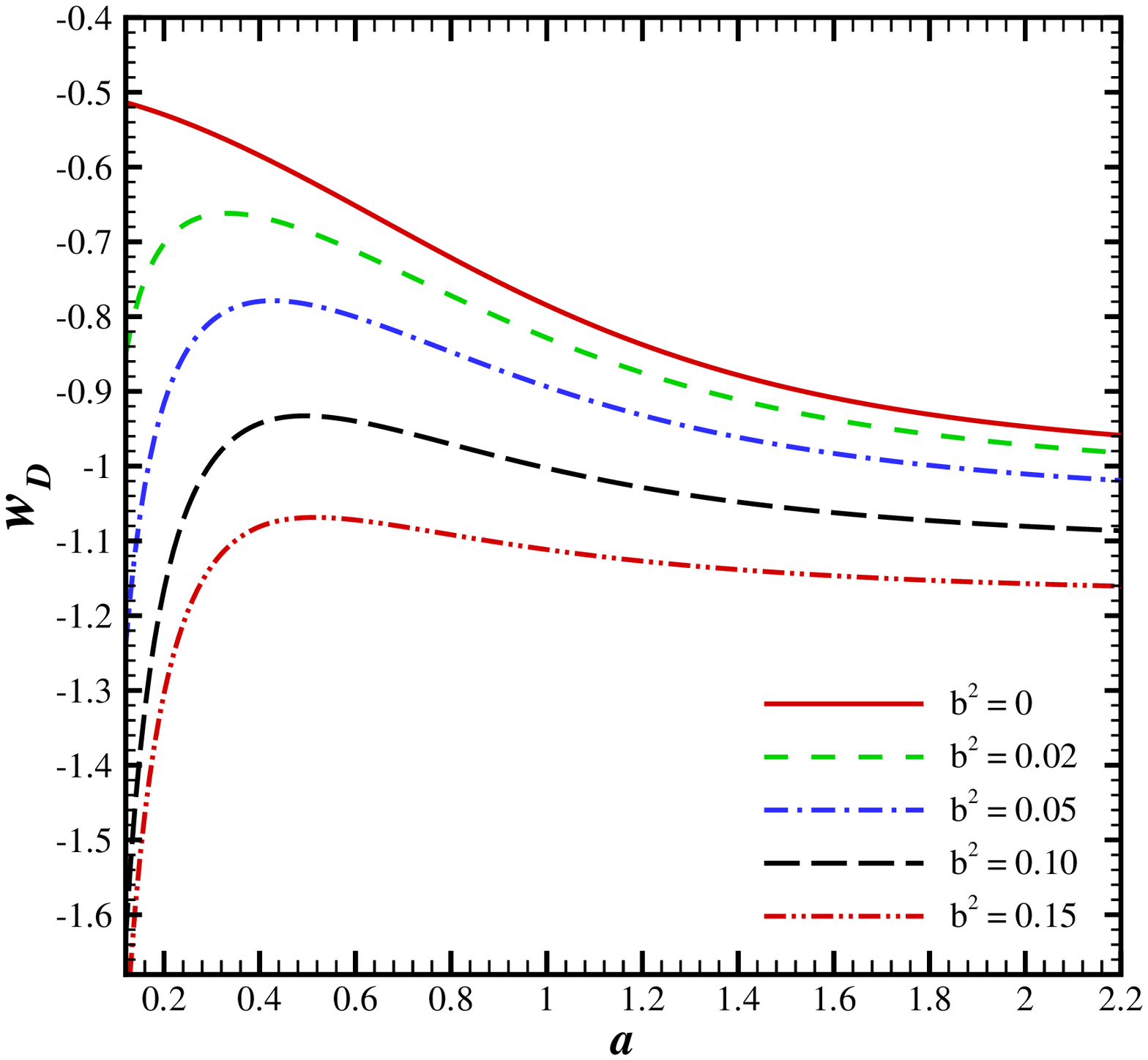}
\includegraphics[width=8cm]{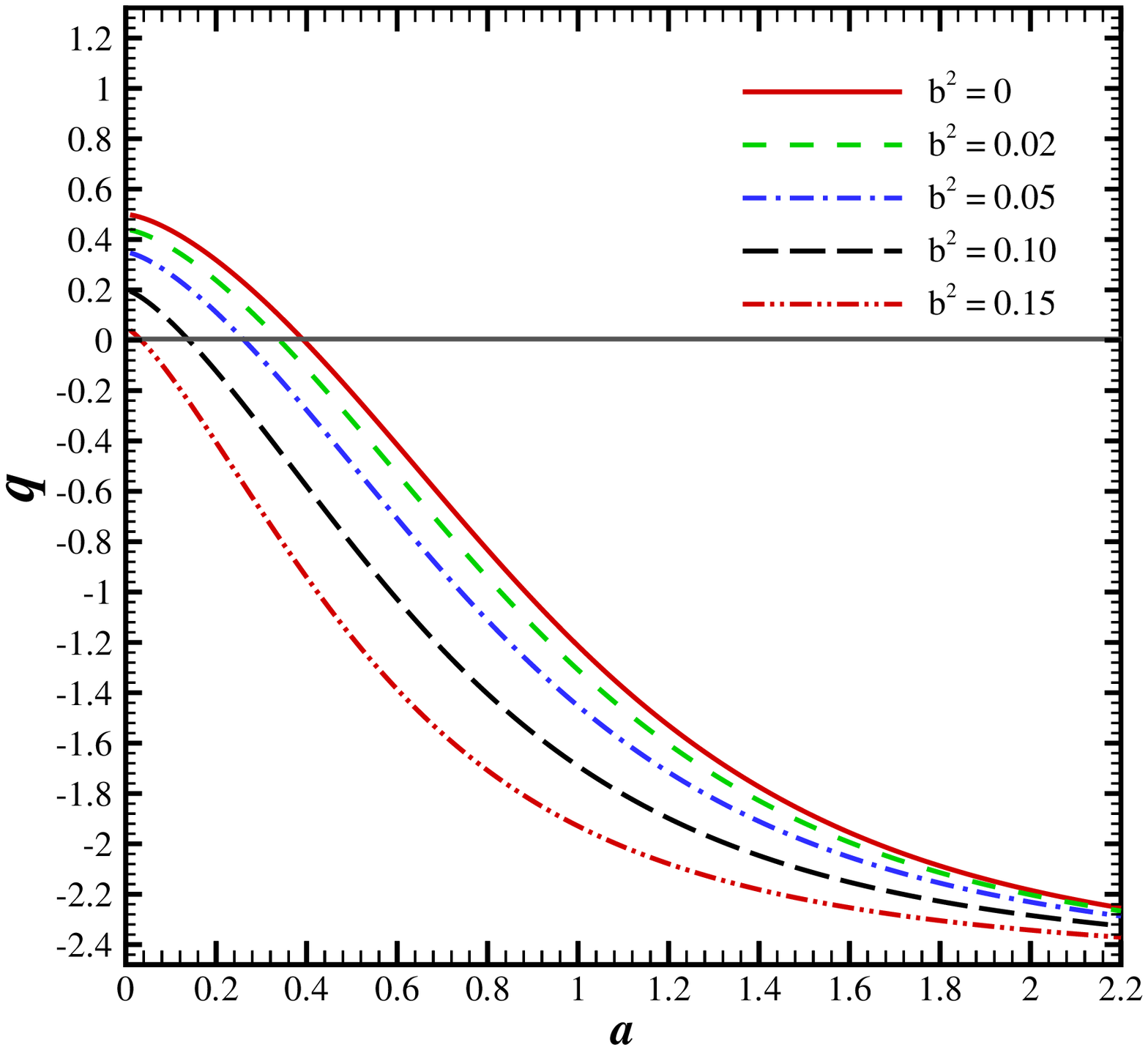}
\caption{The evolution of $w_D$ for interacting ghost dark energy in
nonflat universe. Right panel illustrates the evolution of the
deceleration parameter for interacting ghost dark energy in nonflat
universe. Here we set $\Omega^{0}_{D}=0.73$ and
$\Omega^{0}_{m}=0.28.$ }\label{fig8}
\end{center}
\end{figure}
\begin{figure}[htp]
\begin{center}
\includegraphics[width=8cm]{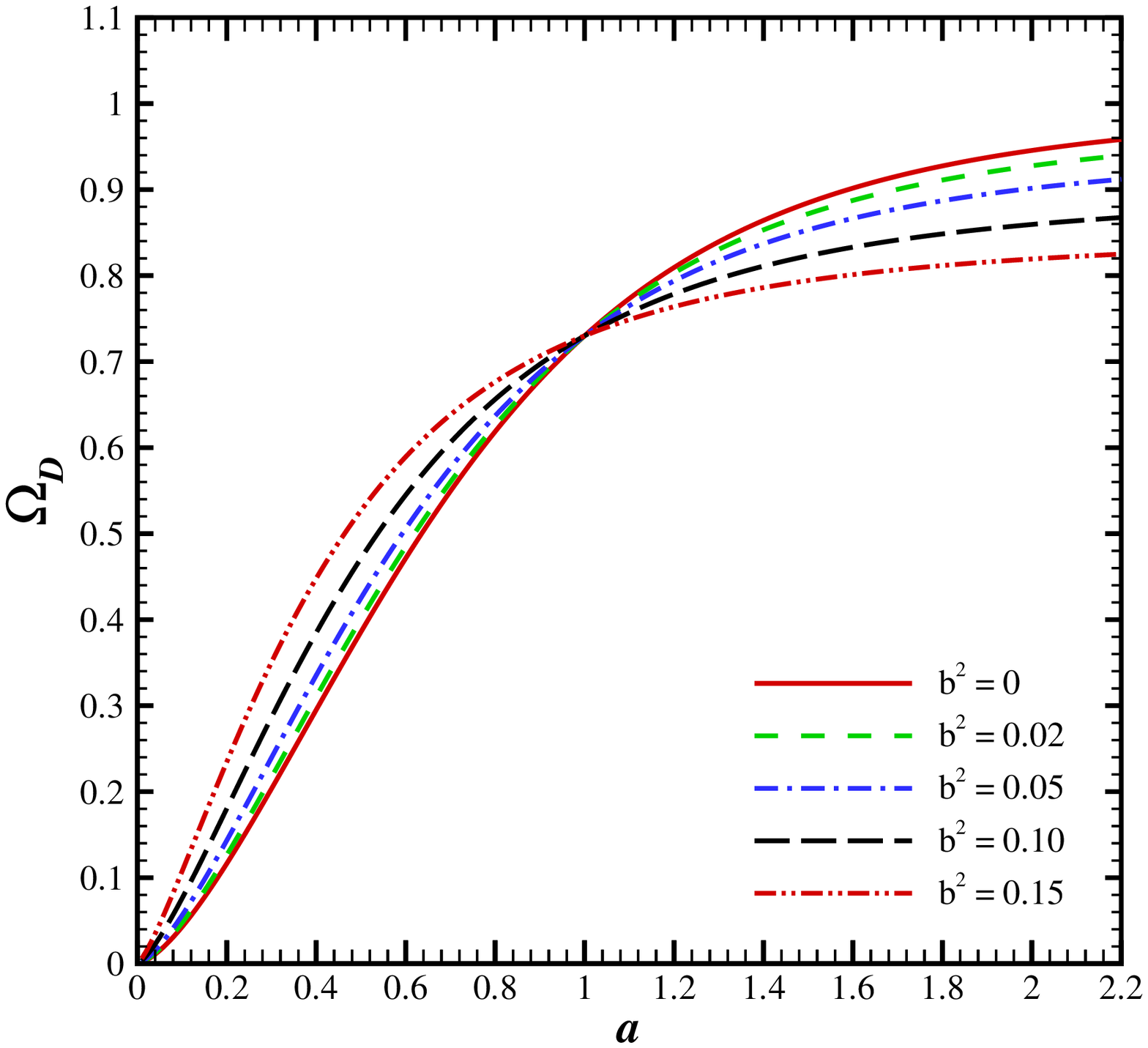}
\includegraphics[width=8cm]{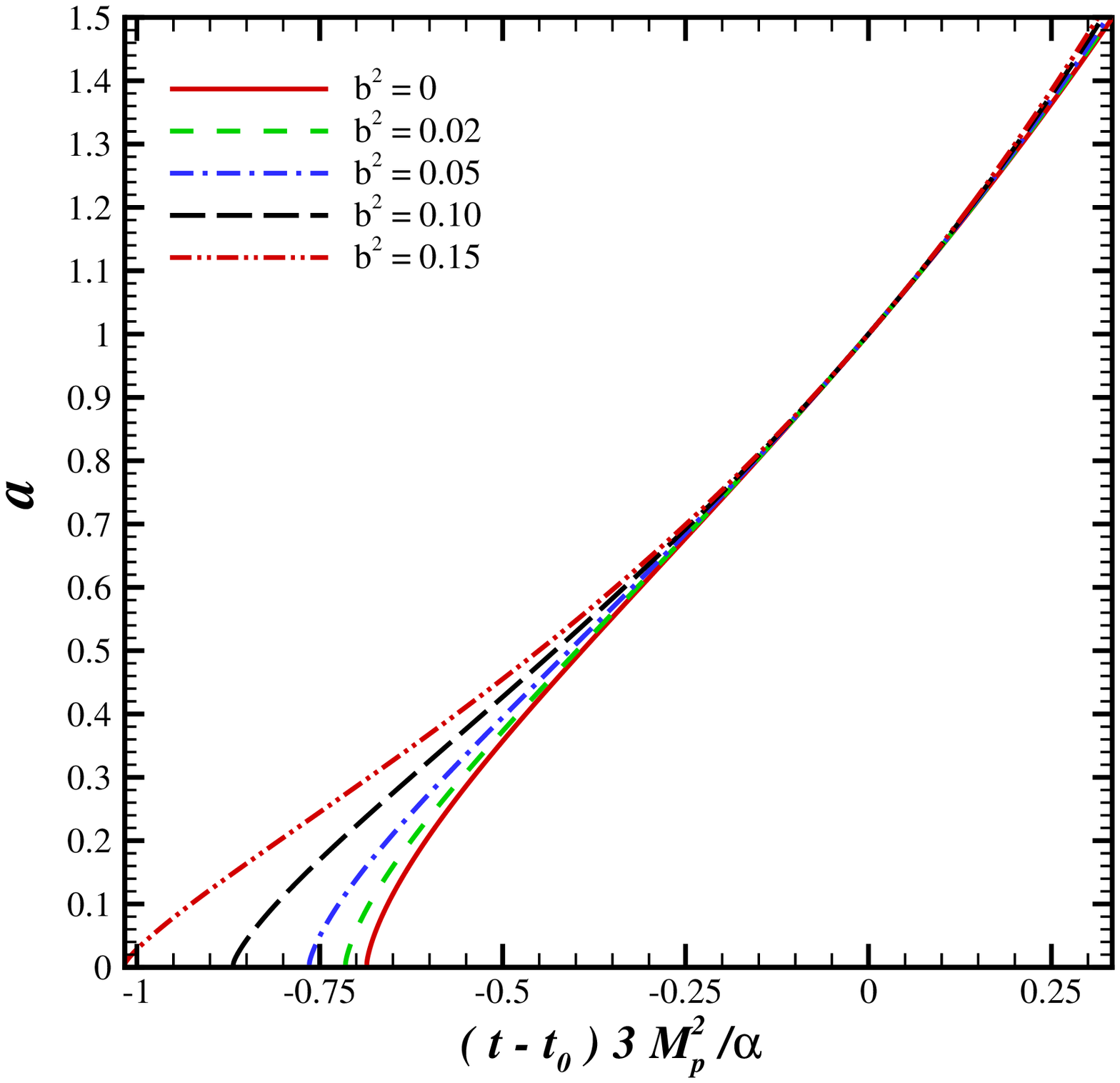}
\caption{Left panel shows the evolution of dark energy density  for
interacting ghost dark energy in nonflat universe. Right panel
corresponds to the evolution of the scale factor for interacting
ghost dark energy in non flat universe. The value of $\Omega_k$ in
the present time is $0.01$ (closed universe). The rest of parameter
are as in figure (\ref{fig8}). }\label{fig10}
\end{center}
\end{figure}

In a non-flat FRW universe, the equation of motion of interacting
ghost dark energy is obtained following the method of the previous
section. The result is
\begin{equation}\label{Omegaprime2n}
\frac{d\Omega_D}{d\ln
a}=\frac{3}{2}\Omega_D\left(1+\frac{\Omega_k}{3}-\frac{\Omega_D}{2-\Omega_D}\left[1-\frac{\Omega_k}{3}+2b^2
\Omega_D^{-1} \left(1+\Omega_k\right)\right]\right).
\end{equation}
The evolution of $\Omega_k$ can be obtained by combining Eq.
(\ref{Omega}) with definition  $\Omega_k=k/(a^2H^2)$. We find
\begin{equation}\label{Omegak}
\Omega_k=\frac{k}{a^2H^2}=\left(\frac{9 M_p^4 k}{\alpha^2}\right)\frac{\Omega_D^2}{a^2}.
\end{equation}
We calculated the evolution of deceleration parameter and $\Omega_D$
and plotted them in figures (\ref{fig8}) and (\ref{fig10}),
respectively. $a(t)$ versus $t$ in the non-flat universe for
different values of coupling constant is shown in figure
(\ref{fig10}). In the limiting case $\Omega_k=0$, Eqs. (32)-(35),
restore their respective equations in interacting ghost dark energy
model in flat universe derived in the previous section (see also
\cite{CaiGhost}).
\section{Observational constraints}
In this section we use the recent observational data sets for
supernova type Ia (SNIa)\cite{Riess04,Riess07}, shift parameter of Cosmic Microwave
Background Radiation based on WMAP-7
\cite{spe03,wmap7,wmap71} and Baryonic Acoustic Oscillation (BAO)
based on Sloan Digital Sky survey (SDSS)\cite{Eis05} to put
constraints on the free parameters of our model. To avoid the
rewriting unnecessary things we refer the reader to some references
such as \cite{movahed06,sheykhi07,baghram08,baghram09} for more
details. In Table \ref{tab1}, we summarize the list of free parameters
of model as well as priors for using in the likelihood analysis.
\begin{table}
\begin{center}
\caption{\label{tab1} Priors on the free parameter space.}
\medskip
\begin{tabular}{|c|c|c|}
  \hline
  {\rm Parameter}& Prior & \\  \hline
  $\Omega_m^0$ & $[0.00-1.00]$  & {\rm Top hat}\\\hline
$\Omega_D^0$ & $[0.00-1.00]$  & {\rm Top hat}\\\hline
 $H_0$&$-$&{\rm Free \cite{hst,zang}}\\\hline
 $b^2$&$[0.00-0.20]$&{\rm Top hat}\\\hline
 \end{tabular}
\end{center}
\end{table}

To apply the observations from SNIa we calculate the distance
modulus as
\begin{eqnarray}
\mu\equiv
m-M&=&5\log{D_{L}(z;\Omega_m^0,\Omega_D^0,b^2)}+5\log{\left(\frac{c/H_0}{1\quad
Mpc}\right)}+25, \label{eq:mMr}
\end{eqnarray}
in the above equation
\begin{eqnarray}
\label{luminosity} D_L (z;\Omega_m^0,\Omega_D^0,b^2) &=&{(1+z) \over
\sqrt{|\Omega_k^0|}}\, {\cal F} \left( \sqrt{|\Omega_k^0|}\int_0^z\,
{dz'H_0\over H(z';\Omega_m^0,\Omega_D^0,b^2)} \right).
\end{eqnarray}
where $${\cal{F}}(x)\equiv(x,\sin(x),\sinh(x))\quad {\rm for}\quad
(\Omega_k^0=0,\Omega_k^0>0,\Omega_k^0<0)$$ and $H(z;\Omega_m^0,\Omega_D^0,b^2)$ is computed
numerically from Eqs. (\ref{Fried}), (\ref{GDE}), (\ref{Omega}) and
(\ref{Omegaprime2n}).

Finally the $\chi^2_{\rm SNIa}$ is defined by:
\begin{eqnarray}\label{chi_sn}
\chi^2_{\rm
SNIa}(\Omega_m^0,\Omega_D^0,b^2)&=&\sum_{i}\frac{[\mu_{obs}(z_i)-\mu_{the}(z_i;\Omega_m^0,\Omega_D^0,b^2)]^2}{\sigma_i^2}
\end{eqnarray}

\begin{figure}[htp]
\begin{center}
\includegraphics[width=8cm]{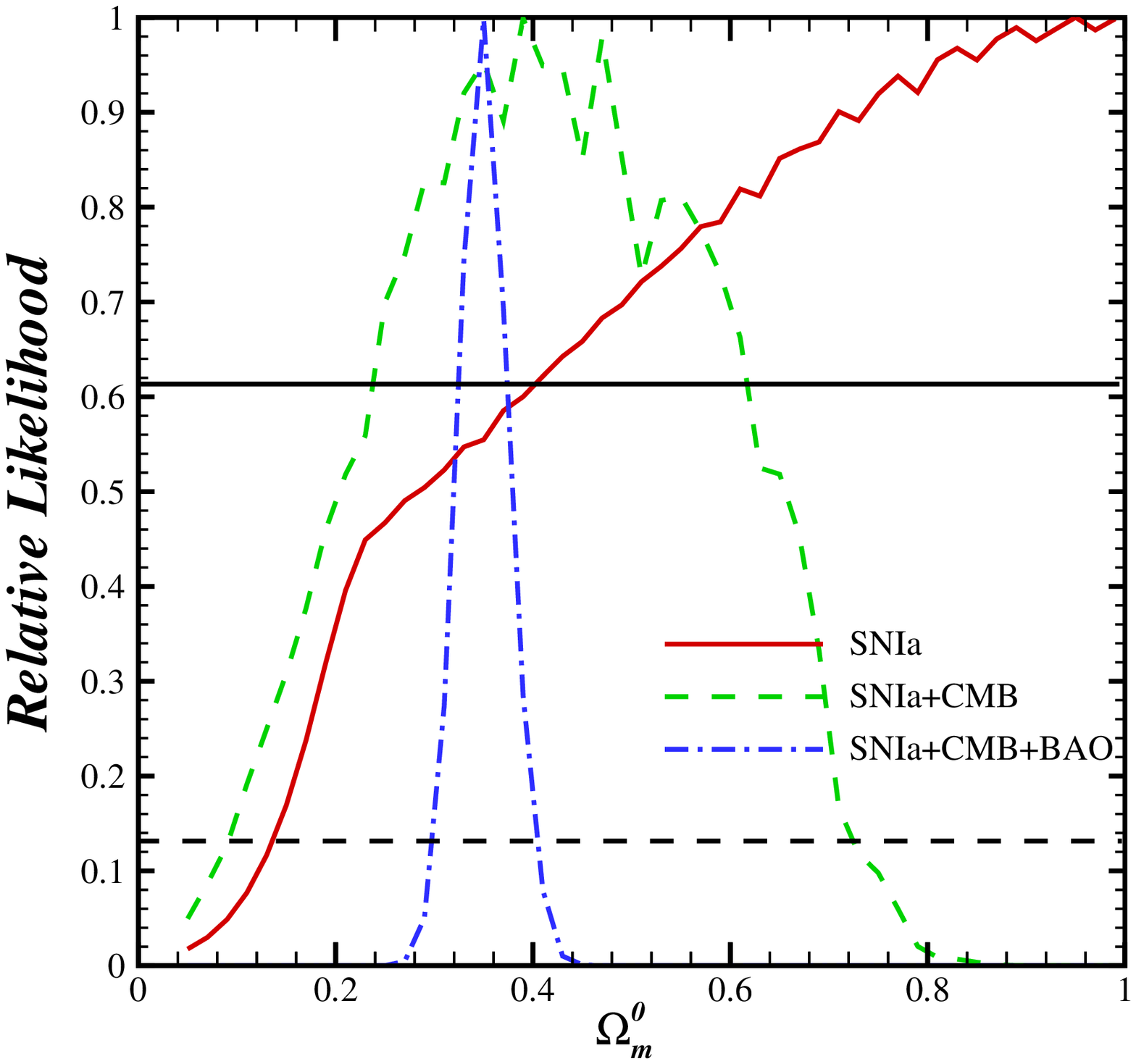}
\includegraphics[width=8cm]{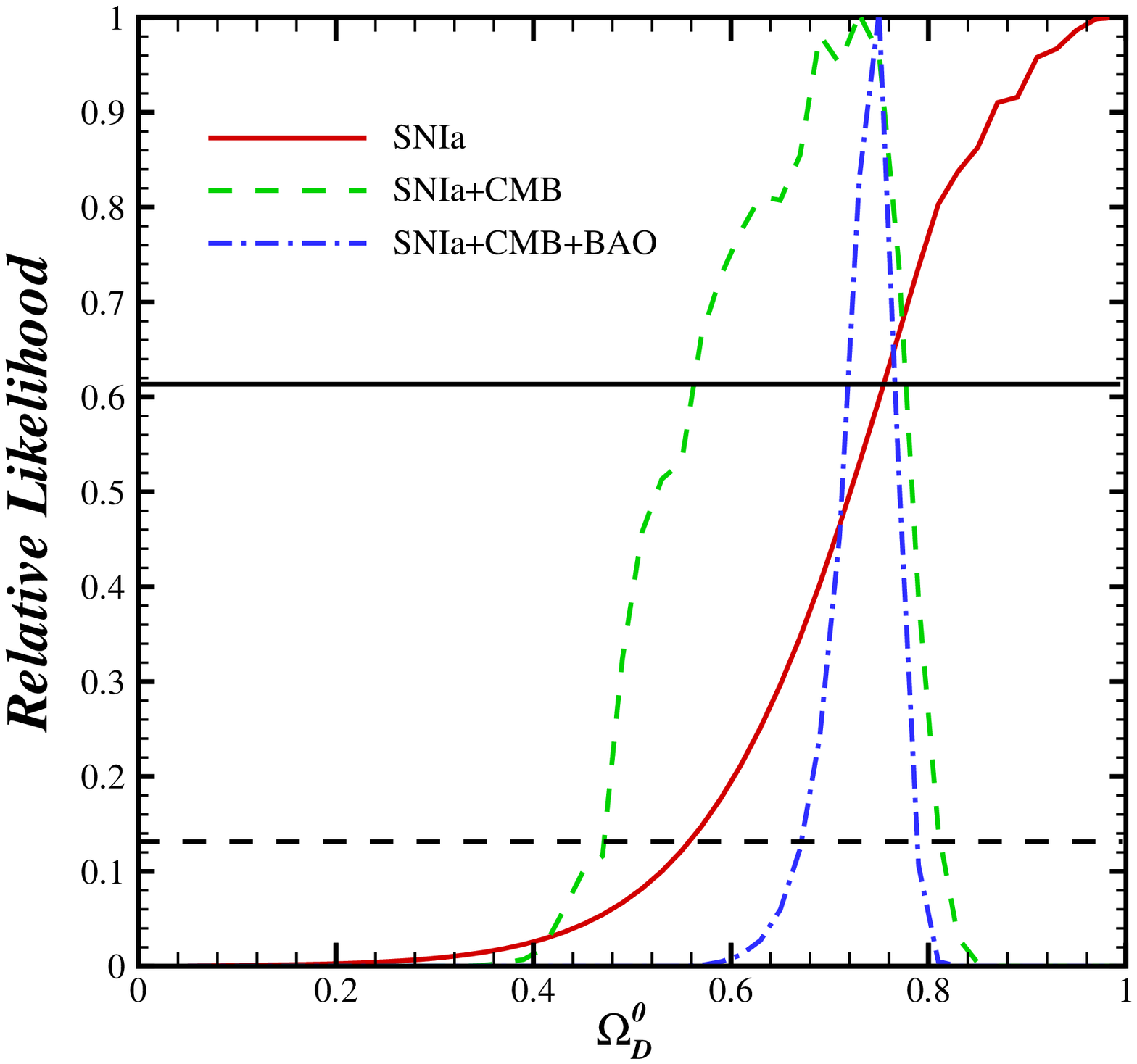}
\includegraphics[width=8cm]{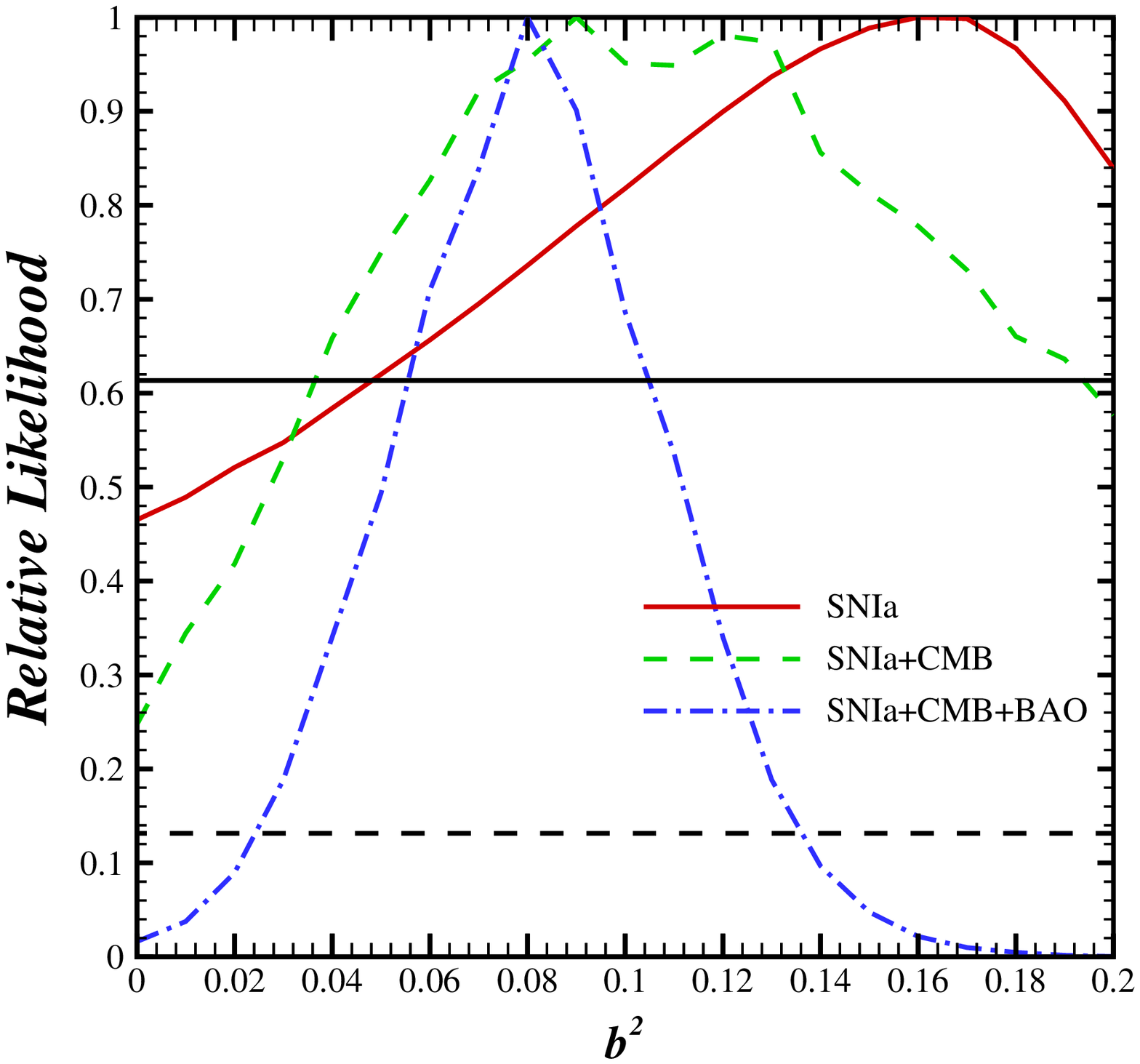}
\caption{Marginalized likelihood functions of model free parameters.
The solid, dash and dashdot lines correspond to fitting the model
with SNIa data new gold sample, SNIa$+$CMB and SNIa$+$CMB$+$BAO,
respectively. The horizontal solid and dashed lines represent the
bounds with $1\sigma$ and $2\sigma$ level of confidence,
respectively.}\label{fig14}
\end{center}
\end{figure}

\begin{figure}[htp]
\begin{center}
\includegraphics[width=8cm]{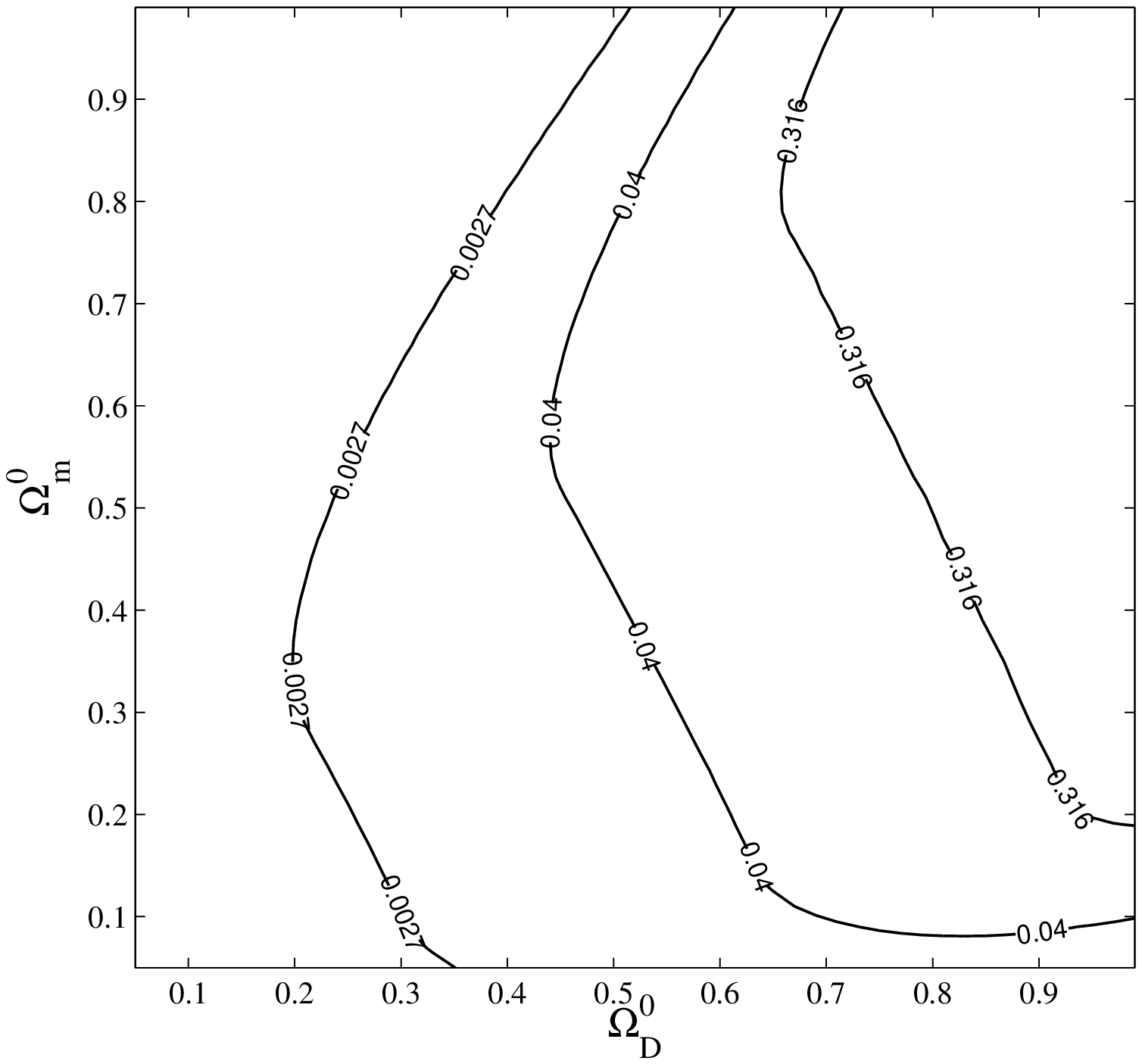}
\includegraphics[width=8cm]{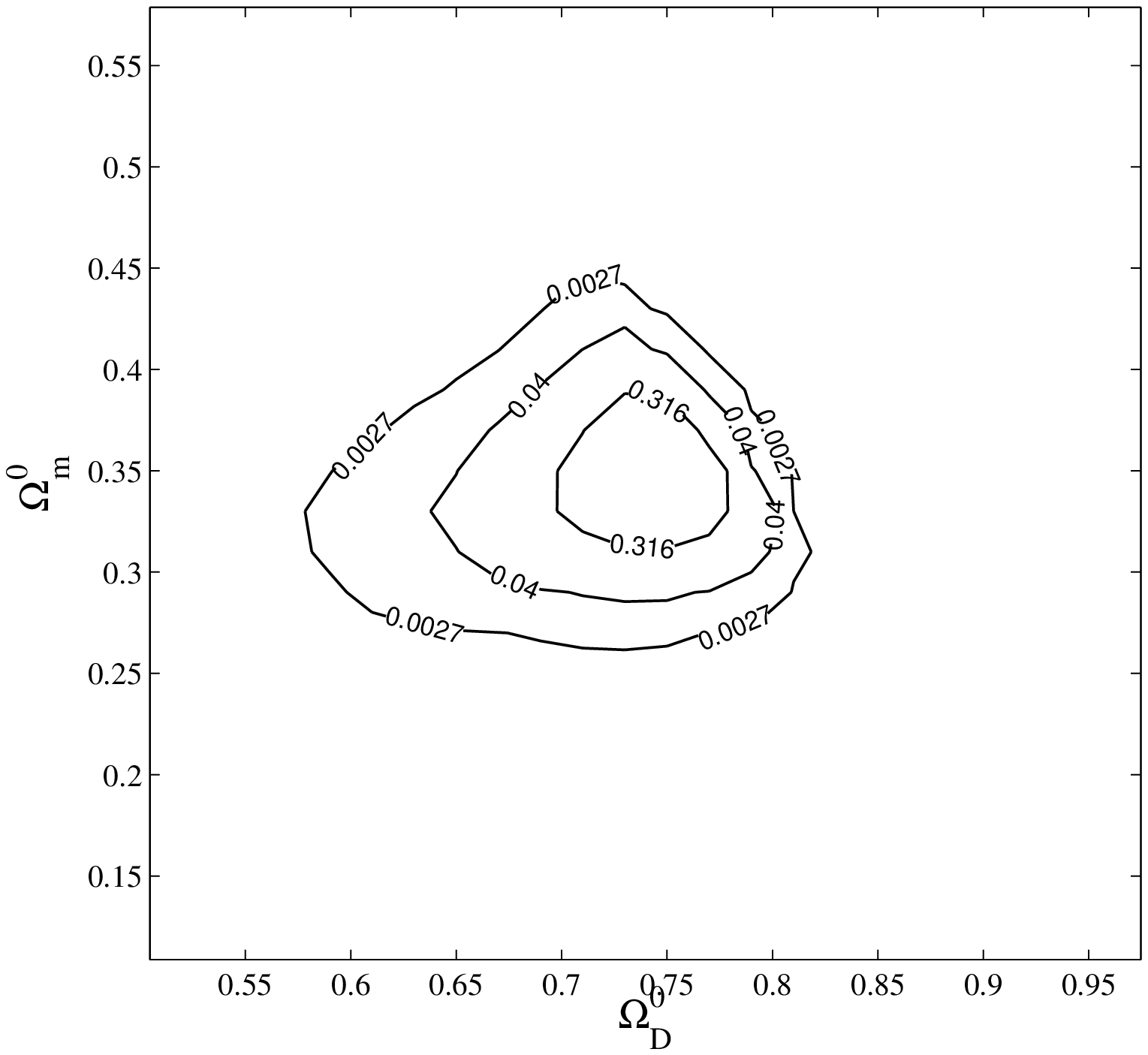}
\caption{Joint likelihood function of $(\Omega_D^0,\Omega_m^0)$. Left
panel corresponds to  SNIa observation while right panel shows
SNIa+CMB+BAO data sets.}\label{fig15}
\end{center}
\end{figure}

\begin{figure}[htp]
\begin{center}
\includegraphics[width=8cm]{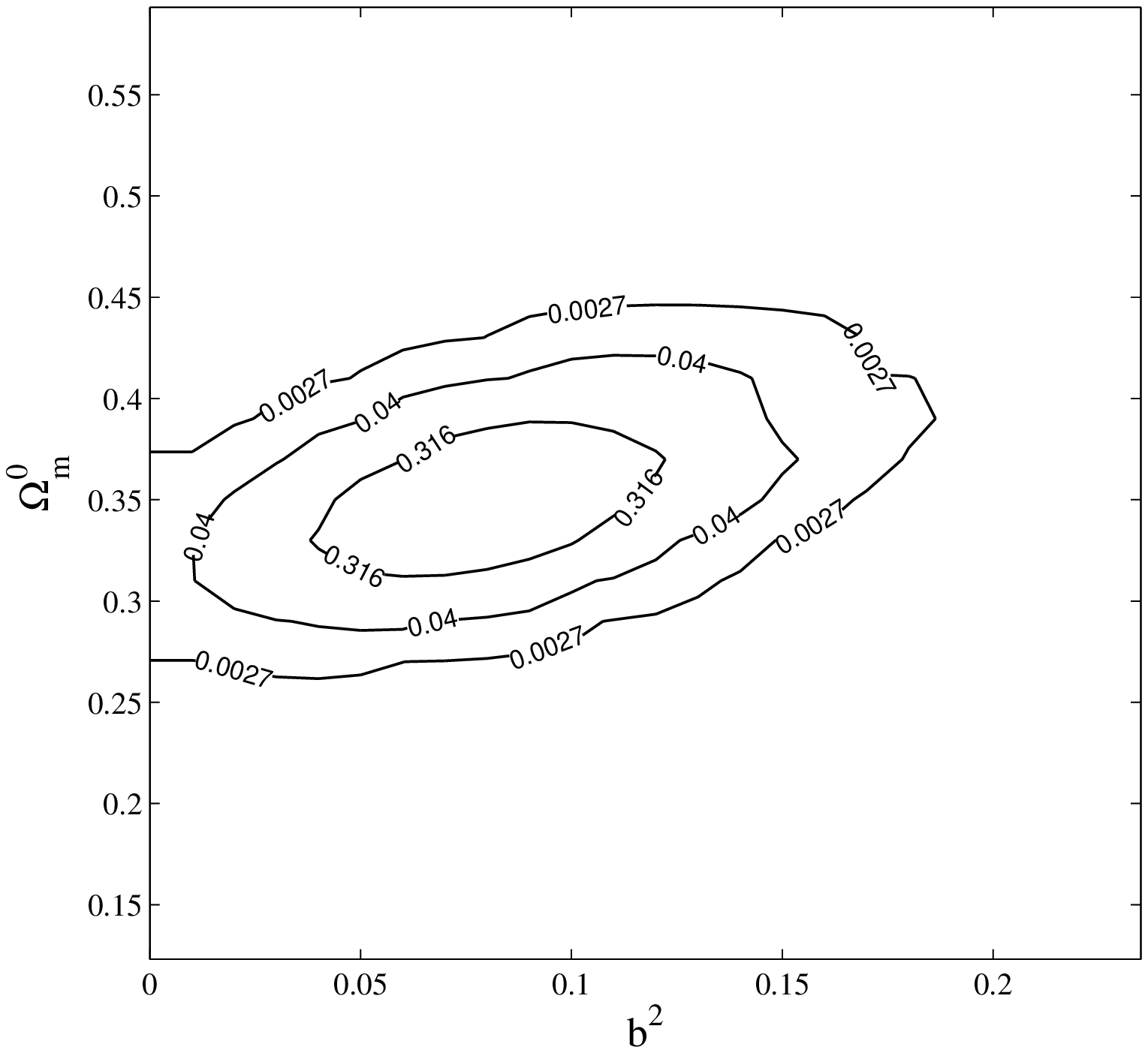}
\includegraphics[width=8cm]{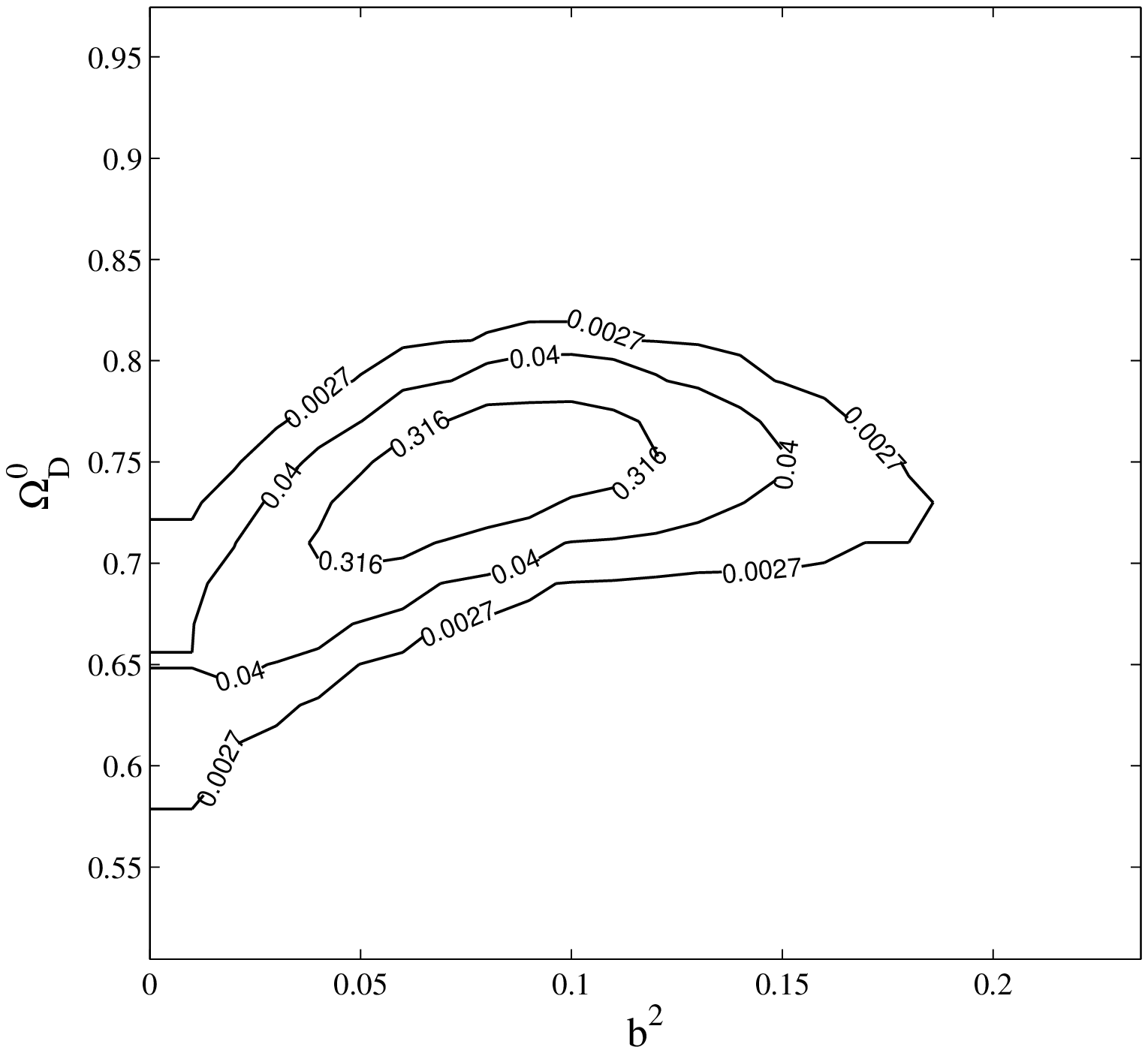}
\caption{Joint likelihood function of free parameters. left panel
corresponds to $(b^2,\Omega_m^0)$ while right panel shows for
$(b^2,\Omega_D^0)$. Here SNIa+CMB+BAO observations used to confine the
values of free parameters.}\label{fig16}
\end{center}
\end{figure}

Usually, beside using the peak locations of the CMB power spectrum,
one can use the so-called shift parameter ${\cal R}$, as \cite{bond97}
\begin{equation}\label{shift_th}
\label{shift} {\cal R}=
\sqrt{\Omega_m^0}\frac{D_L(z_{dec},\Omega_m^0,\Omega_D^0,b^2)}{(1+z_{dec})}
\end{equation}
here $z_{dec}$ is the redshift of the last scattering surface
\cite{Hu95}. Subsequently the $\chi^2_{\rm CMB}$ can be written as
\begin{eqnarray}\label{chi_cmb}
\chi^2_{\rm
CMB}(\Omega_m^0,\Omega_D^0,b^2)&=&\frac{[\mathcal{R}_{obs}-\mathcal{R}_{the}(\Omega_m^0,\Omega_D^0,b^2)]^2}{\sigma^2_{\rm CMB}}
\end{eqnarray}
For the last observational constraint, we rely on the large-scale
correlation function measured from the sample of SDSS including a clear
peak at 100 Mpc$h^{-1}$ \cite{Eis05}. A dimensionless and $H_{0}$
independent parameter for constraining the cosmological models has
been proposed in literatures \cite{Eis05} as follows:
\begin{equation} \label{lss1}
{\cal A} = \sqrt{\Omega_m^0}\left[\frac{H_0D_L^2(z_{\rm
sdss};\Omega_m^0,\Omega_D^0,b^2)}{H(z_{\rm
sdss};\Omega_m^0,\Omega_D^0,b^2)z_{\rm sdss}^2(1+z_{\rm
sdss})^2}\right]^{1/3}
\end{equation}
where $z_{sdss}=0.35$ \cite{Eis05}. So the $\chi^2_{\rm BAO}$ is
expressed as:
\begin{eqnarray}\label{chi_bao}
\chi^2_{\rm
BAO}(\Omega_m^0,\Omega_D^0,b^2)&=&\frac{[\mathcal{A}_{obs}-\mathcal{A}_{the}(\Omega_m^0,\Omega_D^0,b^2)]^2}{\sigma^2_{\rm BAO}}
\end{eqnarray}
Figure (\ref{fig14}) represent the marginalized likelihood function
for model free parameters. In addition joint contour plot for
parameters have been illustrated in figures (\ref{fig15}) and
(\ref{fig16})


The best values and the confidence interval for free parameter at
$1\sigma$ and $2\sigma$ have been reported in Table (\ref{tab2}).

\begin{table}
\begin{center}
\caption{\label{tab2} The best fit values for the free parameters
from fitting with SNIa from new Gold sample, SNIa+CMB, SNIa+CMB+BAO
experiments at one and two $\sigma$ confidence level.}
\begin{tabular}{|c|c|c|c|}
\hline Observation & $\Omega_m^0$ &$\Omega_D^0$ & $b^2$
\\ \hline
  &&& \\
 & $0.95_{-0.56}$& $0.99_{-0.25}$&$0.16_{-0.11}$\\ 
 SNIa &&&\\
 & $0.95_{-0.80}$& $0.99_{-0.44}$&$0.16_{-0.16}$
\\ &&&\\ \hline
&&&\\
& $0.39^{+0.22}_{-0.16}$&$0.73_{-0.17}^{+0.04}$&$0.09^{+0.10}_{-0.05} $  \\
SNIa+CMB&&&\\
 &$0.39^{+0.34}_{-0.30}$&$0.73_{-0.26}^{+0.08}$&$0.09_{-0.09}$ \\
 &&&\\ \hline
 &&&\\
& $0.35^{+0.02}_{-0.03}$&$0.75_{-0.04}^{+0.01}$& $0.08^{+0.03}_{-0.03}$ \\
SNIa+CMB+BAO&&&\\
&$0.35^{+0.06}_{-0.07}$&$0.75_{-0.08}^{+0.04}$&      $0.08^{+0.05}_{-0.05} $\\
&&&\\ \hline
\end{tabular}
\end{center}
\end{table}

\section{Conclusion}\label{sum}
It is a general belief that our universe is currently undergoing a
phase of  accelerated expansion likely driven by dark energy.
Unfortunately, until now, the nature and the origin of such dark
energy is still the source of much debate and  we don't know what
might be the best candidate for dark energy to explain the
accelerated expansion. Thus, various models of dark energy have been proposed,
to explain the accelerated expansion by
introducing new degree of freedom or by modifying the standard
model of cosmology. In this regard, a so called ``ghost dark energy"
was  recently proposed \cite{Urban,Ohta} which originates from the Veneziano ghost of QCD.
The QCD ghost has no contribution to the
vacuum energy density in Minkowski spacetime, but in curved
spacetime it gives rise to a small vacuum energy density \cite{Ohta}.
The dark energy density is proportional to Hubble parameter,
$\rho_D=\alpha H$, where $\alpha$ is a constant of order $\Lambda_{\rm QCD}^3$ and
$\Lambda_{\rm QCD}$ is QCD mass scale. With $\Lambda_{\rm QCD}\sim
100MeV$ and $H\sim 10^{-33}eV$ ,  $\Lambda_{\rm QCD}^3 H$ gives
the right order of magnitude $\sim (3\times 10^{-3}\rm {eV})^4$
for the observed dark energy density \cite{Ohta}. The
advantages of this new proposal compared to the previous dark
energy models is that it totally embedded in standard model so
that one needs not to introduce any new parameter, new degree of
freedom or to modify general relativity \cite{CaiGhost}.

In this paper, we generalized the ghost dark energy model, in the
presence of interaction between dark energy and dark matter, to the
universe with spatial curvature. Although it is believed that our
universe is spatially flat, a contribution to the Friedmann equation
from spatial curvature is still possible if the number of e-folding
is not very large \cite{Huang}. Besides, some experimental data has
implied that our universe is not a perfectly flat universe and
recent papers have favored the universe with spatial curvature
\cite{spe}. With the interaction between the two different dark
components of the
 universe, we studied the evolution of the universe,
from early deceleration to late time acceleration. We found that in
the absence of interaction the equation of state parameter of ghost
dark energy is always larger than $-1 $ and mimics a cosmological
constant in the late time. We also found that the transition from
deceleration to acceleration take places at $a\simeq0.64$ or
equivalently at redshift $z\simeq 0.56$. We observed that, in the
presence of interaction, the equation of state parameter can cross
$-1$ at the present time provided the interacting parameter satisfy
$b^2>0.1$.

 To check the observational consistency of
interacting Ghost Dark Energy model, we used Supernova type Ia
(SNIa), CMB shift parameter and Baryonic Acoustic Oscillation (BAO).
Our results demonstrated that the best values of free parameters
when we combine all observational data are: $\Omega_m^0=
0.35^{+0.02}_{-0.03}$, $\Omega_D^0=0.75_{-0.04}^{+0.01}$ and
$b^2=0.08^{+0.03}_{-0.03}$ at $1\sigma$ confidence interval. Our
analysis shows that at $1\sigma$ level of confidence the value of
so-called interacting parameter does not cross zero. Also the total
value of energy density of universe at present time is $\Omega_{\rm
tot}^0=\Omega_m^0+\Omega_D^0=1.10^{+0.02}_{-0.05}$  at $68\%$ level.

Finally, we would like to mention that if there is any kind of
ghost field which gives rise to an energy density $\rho \propto
H$, its cosmological implications is exactly similar to the
present work independent of its origin. Although the existence of
a well-motivated physical model where this energy density is
obtained is of course a valid starting point, however, the details
of this study can be found in the previous works such as
\cite{Urban,Ohta} and we have not repeated them here. In this work
our main task was to study the cosmological implications of this
new ghost dark energy model proposed in \cite{Urban,Ohta} without
referring to its origin. In particular we generalized the study to
the universe with spatial curvature and put some observational
constraints on the model parameters.

\acknowledgments{A. Sheykhi thanks Professor Rong Gen Cai for
valuable comments and useful discussion. This work has been
supported by Research Institute for Astronomy and Astrophysics of
Maragha, Iran.}

\end{document}